\documentstyle[11pt]{article}
\input{psfig}
\begin{document}
\begin{flushright}
Liverpool Preprint: LTH 342\\
hep-lat/9412087\\
Dec 19, 1994
\end{flushright}
\vspace{5mm}
\begin{center}
{\LARGE\bf
Fitting Correlated Hadron Mass Spectrum Data
}\\[1cm]

{\bf C. Michael and A. McKerrell}\\

\it{DAMTP, University of Liverpool, Liverpool, L69 3BX, U.K.}

\end{center}

\begin{abstract} 

We discuss fitting hadronic Green functions versus time $t$ to  extract
mass values in quenched lattice QCD.  These data are themselves strongly
correlated  in $t$. With only a limited number of data  samples,  the
method of minimising correlated $\chi^2$ is unreliable. We explore 
several methods of modelling the correlations among the data set  by a
few parameters which then give a stable and sensible fit even if the
data sample is  small. In particular these models give a reliable 
estimate of the goodness of fit.

 \end{abstract}

\section{Introduction}

We assume that we have $N$ samples of unbiased estimators of  quantities
$x(t)$ with $t=1 \dots D$. Thus the data set is  $x^{(n)}(t)$ where $n=1
\dots N$. We assume that the samples $x^{(n)}(t)$ are statistically
independent versus $n$ for  fixed $t$ but may be correlated in $t$. 
Such a situation arises in lattice gauge theory calculations  where
there are $N$ independent configurations and $D$  Green functions
(vacuum expectation values of  combinations of quark  propagators and/or
gauge links) are measured versus time separation $t$. An  introduction
to this topic in the context of  lattice gauge theory is provided by
Toussaint~\cite{T}.

The aim is to fit a given function $F(t)$ which depends on $P$  
parameters $a_p$. This function is to be fitted to the 
data samples $x(t)$.   Thus we require to find the following

\begin{itemize}
\item the best values of the parameters $a_p$. 
\item the errors associated with these best fit parameters.
\item the confidence level that the fit represents the data sample.
\end{itemize}

A general discussion of this problem has been given~\cite{cmfit}  which
has shown that with limited data samples $N$,  it may be unrealistic to
allow a general form for the $D \times D$  correlation matrix describing
the correlations among the  data at the $D$ different $t$ values. In
particular the  estimated $\chi^2$ was shown to be increased by a factor
of $1+(D+1)/N + {\cal O}(N^{-2})$. This led to the  conclusion  that  
$N > 10(D+1)$  is needed for a reliable use of $\chi^2$ as a  goodness of
fit estimator in a correlated $\chi^2$ fit.

Although this conclusion arose from studying a theoretical  distribution
which was uncorrelated, the above $\chi^2$ increase is the  same for any
true distribution. Indeed the $\chi^2$ per degree of freedom will have
the expected value $(N-1)/(N-D-2)$. Furthermore the distribution of 
$\chi^2$ can be evaluated theoretically~\cite{and}  so that confidence
levels can  be derived.  The major consequence of small sample size is
that  the eigenvalues of the correlation matrix are modified: in
particular  very small eigenvalues can arise. These correspond to a very
narrow distribution in the corresponding  eigen-direction and so can
bias any fit considerably.

In order to cope with this, it is necessary to use some extra 
theoretical input. In particular, one needs to assume a form of  the
correlation matrix or of its eigenvalue spectrum. This has  been
recognised before, and proposals have been made to truncate  the
eigenvalues according to the spirit of the SVD inverse of a singular 
matrix~\cite{leth,drho,kilcup}. We discuss this proposal and offer  our
suggestions for an improved treatment. We believe that our  approach is
more stable.

Another way to attack this problem is to explore the theoretical 
expectations for the correlation matrix. In the case we are considering,
the correlation between meson operators can be expressed in terms of 
vacuum expectation values of 4-quark operators~\cite{leth}.  These can
be estimated  and this gives a motivation for a direct parametrization
of the correlation  matrix in terms of exponentials in $|t-t'|$. If the
normalised correlation is given by one such exponential, then its
inverse is a tridiagonal matrix.  This is a very appealing model since
it corresponds to a nearest neighbour  linkage in the underlying
probability distribution. This one-parameter model is very stable, but
does not always give an accurate  description of the data. A natural way
to extend it is to consider  a 5-diagonal inverse as an efficient
parametrization. This 2-parameter model corresponds to a correlation
matrix given by a specific combination  of 2 exponentials and it fits
the data well in many cases.

We start by presenting typical data on the hadronic Green functions
from quenched lattice studies. This enables us to analyse the 
behaviour of the correlation matrix which is at the heart of 
our exploration. We then compare various methods of modelling 
the correlation matrix and test their stability in correlated 
$\chi^2$ fits.

\section{Fits with  correlated  data }

The data $x(t)$ we will consider  are Green functions which are vacuum
expectations of hadronic operators at times 0 and $t$. Such  data are
often referred to as hadron correlators but we will not use this 
description here since we wish to concentrate on a different
correlation: that  between the Green functions $x(t)$ at different $t$
values. 

The data sample themselves give a probability distribution
$$
S(x)={1 \over N} \sum_{n=1}^{N} \delta^D (x-x^{(n)})
$$
 We shall be interested in estimates of the probability distribution 
$Q(X)$ of the averages $X(t)$ of the data $x(t)$. A smooth
representation  of $Q(X)$ is needed  for determining best fit parameters
and to estimate the acceptability  of such a  fit. The natural
parametrisation  for $Q$ is suggested by the central limit  theorem.
Provided that the underlying distributions of $x(t)$ are  sufficiently
localised, then for large $N$, $X(t)$ will be  gaussianly distributed. 
We are specifically interested in the  case where the different
components $x(t)$ are statistically  correlated. Thus a general gaussian
surface will be needed. 
 $$ Q(X)= H \exp( -{1
\over 2} \sum_{t,t'} (X(t) - \overline{X}(t)) M(t, t') (X(t') -
\overline{X}(t') ) ) 
 $$
 with
 $$ 
\overline{X}(t)={1 \over N} \sum_{n=1}^N x^{(n)}(t)
$$
$$
M(t,t')=N C^{-1}(t,t') \ \hbox{ ,where} 
$$
$$
C(t,t')=    {1 \over N-1} \sum_{n=1}^N 
(x^{(n)}(t)-\overline{X}(t))(x^{(n)}(t')-\overline{X}(t'))
$$
To find the best fit parameters then corresponds to maximising
$$
\exp(-\chi^2 /2) \ \hbox {where } \
 \chi^2= \sum_{t,t'} 
   (F(t,a)-\overline{X}(t)) M(t,t') (F(t',a)-\overline{X}(t')) 
 $$
 with respect to $a_p$ for $p=1 \dots P$. This is the usual  correlated
$\chi^2$ method. For sufficiently large $N$, this is a  stable procedure
and the expected value of $\chi^2$ is the number of  degrees of freedom
$D-P$. But for small $N$, the sampling fluctuation in $C$ and  hence $M$
can give unreasonable fits. In order to avoid this bias, we  aim to make
a more stable model for $C$.

Let us now study the properties of the correlation matrix $C$. It is a 
real symmetric positive-definite matrix for any number of samples $N$
but it has rank $N-1$ so will have $D-N+1$ zero eigenvalues if $N \le
D$. In this latter case, its inverse is not defined. A commonly used 
prescription in such a case is the Singular Value Decomposition (SVD)
inverse which corresponds to omitting the eigenmodes corresponding to
the zero eigenvectors from the inverse. We will return to discuss the
utility  of this prescription.

For many purposes, it is simpler to study the normalised correlation 
matrix which we define as 
$$
 \tilde{C}(t,t') = { C(t,t') \over \sqrt{C(t,t) C(t',t')} } 
$$

\section{Correlations among hadronic operators}

In quenched lattice determinations of hadronic spectra and matrix 
elements, one studies vacuum expectation values of hadronic  operators
at times 0 and $t$. Thus $x(t)$ is the vacuum expectation value of 
hadronic  operators $<\! H(0)H(t) \! >$.  These determinations of $x(t)$
are extracted from the quark  propagators derived from inverting the
fermion matrix in the  given gauge field sample. This propagator
inversion is very  demanding computationally and is usually only
evaluated for  one source point $({\bf 0},0)$. This implies that   quark
propagators to all values of $t$ are equally sensitive to the region
around this fixed  source site. Thus between different gauge (vacuum)
samples, all  propagators will tend to be large/small as this fixed
region  is conducive/resistant to quark propagation. This  argument
shows that very prominent correlations are expected between  hadron
Green functions to different $t$ values. We analyse some  data to
substantiate this.

We present results  for  the normalised correlation $\tilde{C}(t,t')$ in
fig~1. Here the data come from a study  of the $\pi$ and $\rho$ mesons
using local hadronic operators from a point  source at $({\bf 0},0)$ to
$({\bf y},t)$ summed over ${\bf y}$ to give a  zero-momentum observable.
There are $N=60$ independent configurations using a  hopping parameter
$K=0.14262$ with the clover improved action~\cite{ukqcd}. The lattice
has size $24^3 \times 48$ at $\beta=6.2$  and, for orientation, $m_{\pi}a
\approx 0.17$ at this hopping parameter value. We use $t$ values 5 to 24 
for this study.

We find that ${\tilde C}$ decreases with $|t-t'|$ but is relatively 
insensitive to $t+t'$.  The decrease with $|t-t'|$ is illustrated 
in fig~1.  An exponential behaviour versus $|t-t'|$ is 
expected from an analysis in terms of hadronic operators.

Let us  summarise this argument. Consider the case where $x(t)$ is the
vacuum expectation value of  hadronic  operators $<\! H(0)H(t) \! >$. Then
 $$
 C(t,t') = \,<\! H(0)H(0)H(t)H(t') \! > - <\! H(0)H(t) \! ><\! H(0)H(t') \! >
 $$
The first term then will have contributions between $t$ and $t'$  from 
intermediate states of lowest energy $m$ with the quantum numbers  created by
$H$, the same as those in $x(t)$ itself. 
Between 0 and $t$ (for $t' > t$), the intermediate state 
will have the quantum numbers of $H(0)H(0)$  and so may have a lower energy 
which we write as $2M$.
Ignoring for the moment the disconnected term, and assuming 
that one state only dominates in each case, then 
$$
 \tilde{C}(t,t') = e^{-(m-M)|t-t'|}
$$ 

For the case of the $\rho$ for example: $m=m_{\rho}$ and $M=m_{\pi}$
since a 2$\pi$ state can couple to $HH$ (ie to $\rho \rho$). Thus a
small  exponential rate of decrease is to be expected with exponent
$m_{\rho}-m_{\pi}$. In this case the disconnected term in $C$ will be 
relatively unimportant since it decreases faster than the connected term
 by $\exp(-2(m-M)t)$. The contribution of excited intermediate states
will modify this simple  exponential behaviour except at large $|t-t'|$
where the lowest state dominates. The curve corresponding to the 
ground state exponential $\exp(-(m_{\rho}-m_{\pi})|t-t'|)$ is shown in fig.1,
where it is seen to be a reasonable guide to the large $|t-t'|$ 
behaviour.

For the case where 
$x$ is a pion observable, then both $m$ and $M$ are $m_{\pi}$ and the 
correlation $\tilde{C}(t,t')$ would be constant versus $|t-t'|$. In this 
case the disconnected term will be relatively important. 
The disconnected part will reduce the magnitude of $\tilde{C}$ 
especially when relatively more disconnected terms are present -
such as with a source summed over spatial position. When the source
is at a fixed lattice position $({\bf 0},0)$, as above, then the 
disconnected parts will cancel less completely and we expect $\tilde{C}$
to remain large for large $|t-t'|$.  Indeed it is larger for $\pi$ 
than $\rho$ correlations as shown in fig.~1.

For baryon spectra, the nucleon is the lowest lying 3 quark state and 
so $\tilde{C}$ will be like the pion case above, while the $\Delta$ will 
behave analogously to the $\rho$ above.

\section{Exponential models for correlations among the data}

Since there is some theoretical justification for an exponential 
decrease of $\tilde{C}(t,t')$ with $|t-t'|$, we consider first a simple 
and robust model with just one exponential. Since $\tilde{C}(t,t')$ is
normalised, this results in a one parameter model with parameter $a$ 

 $$
\tilde{C}(t,t') = e^{-a |t-t'|} 
 $$ 

In practice the behaviour of $\tilde{C}(t,t')$ is not exactly
exponential, so one must  choose a reference value of $|t-t'|=t_a$ to
determine a suitable value of $a$. We have found that using $t_a=4$ is a
good choice. Then $a$ is determined by averaging the $D-t_a$ values  of
$\tilde{C}(t,t+t_a)$ obtained from the sample data. This averaging along
the off-diagonal of $\tilde{C}$ also helps to reduce the sampling
fluctuations.

The test of the suitability of a model of $\tilde{C}(t,t')$ for our
purposes is  that its inverse $\tilde{M}$ is stable under fluctuations
in the data  sample used to model $\tilde{C}(t,t')$.  Thus it is
appropriate to study the inverse of  $\tilde{C}(t,t')$. For the case of
a single exponential, the inverse is particularly  simple: it is
tridiagonal. It is given exactly in terms of $\alpha\equiv\exp(-a)$  by  
 $$
\tilde{M}(t,t)= { 1+{\alpha}^2 \over 1-{\alpha}^2}  ,\qquad
\tilde{M}(t,t\pm 1)= { - \alpha \over 1-{\alpha}^2} , \qquad
\tilde{M}(t_m,t_m)= { 1 \over 1-{\alpha}^2} 
 $$
where $t_m$ is the maximum or minimum $t$ value in the matrix 
being inverted.

Thus as $a \to 0$, the elements of $\tilde{M}(t,t')$ increase as
$a^{-1}$. In the limit of large D, one can   estimate the smallest
eigenvalue of $\tilde{C}(t,t')$ (largest of $\tilde{M}(t,t')$) which  is
$a/2$  ($2/a$ respectively) as $a \to 0$. Thus if $a$ is reasonably well
determined  by the correlated data sample, then the value of
$\tilde{M}(t,t')$ will be stable  under sample fluctuations. Thus the
resultant fits will be stable  too. This method provides a stable one
parameter model for  $\tilde{C}(t,t')$. One drawback of the model is
that it does not reproduce very  accurately any non-exponential
behaviour of $\tilde{C}(t,t')$. This can be  taken into account in a
straightforward way by considering  2-exponential models for
$\tilde{C}(t,t')$.

Rather than consider an arbitrary 2-exponential model, we generalise the
tri-diagonal feature of $\tilde{M}(t,t')$ and look for 5-diagonal models
instead.  The algebra is now somewhat messier, but the conclusion is
that a  5-diagonal model for $\tilde{M}(t,t')$ corresponds to a
particular 2-exponential  model for $\tilde{C}(t,t')$ with:
 $$
 \tilde{C}(t,t')={  (1-{\alpha_2}^2) {\alpha_1}^{|t-t'|+1}
                  - (1-{\alpha_1}^2) {\alpha_2}^{|t-t'|+1} 
  \over (\alpha_1-\alpha_2) (1+{\alpha_1} {\alpha_2})   }
 $$  
where to have a sensible interpretation we require 
$ 0 \le Re{\,\alpha_1}, Re{\,\alpha_2} < 1$.

If $\alpha_1$ and $\alpha_2$ are  complex, they must be complex
conjugates. In this case the behaviour  of $\tilde{C}(t,t')$ versus
$|t-t'|$ will be a damped oscillation. Although such a  behaviour is not
strictly excluded, it seems unreasonable to have  anti-correlation at
larger $t$ values so we choose not to allow  that possibility.
Henceforward, we take $\alpha_1$ and $\alpha_2$ to be real. 

Thus we obtain $\alpha_1$ and $\alpha_2$ by comparing the above
expression for $\tilde{C}(t,t')$ with the sample data. The two
parameters can, for instance, be  determined by making a least squares
fit to $\tilde{C}(t,t')$ at $|t-t'|=1$ and $t_a$. A  fit is needed
because in some cases the data may not be  reproducible exactly by the
expression. We find that this 2-parameter assignment to the sample
correlation is stable when inverted for  use in fits. This is plausible
since a 5-diagonal form of the  inverse avoids very small eigenvalues of
 $\tilde{C}(t,t')$.


We give the explicit formula for the inverse of the 5-diagonal 
matrix in the Appendix.

\section{Eigenvalue smoothing of the correlation matrix}

The essence of the problem is that sample values of $\tilde{C}(t,t')$
may have  very small eigenvalues and these influence unreasonably the
inverse  $\tilde{M}(t,t')$ used in modelling the distribution of the
data. An obvious way  to proceed is to modify these unreasonable
eigenvalues by hand. One suggestion~\cite{leth} is to remove the {\it
largest} eigenvalues  of $\tilde{C}(t,t')$ since they will have least
influence on $\tilde{M}(t,t')$. This seems hard  to justify and later
suggestions~\cite{drho,kilcup} have been to remove the {\it smallest}
eigenvalues of $\tilde{C}(t,t')$. This latter suggestion is in the 
spirit of the SVD inverse of a singular matrix: only the contributions 
from the non-zero eigenvalues are retained in the inverse. This 
eigenvalue truncation is  clearly a rather brutal approximation to
$\tilde{M}(t,t')$: its largest components are  being removed. Indeed the
gaussian surface  modelling the probability distribution will be
unconstrained in  the direction of the deleted eigenmodes.  A physical
argument, for why this may be   acceptable in practice, can be based  on
the observation~\cite{drho} that the smallest eigenvalues of
$\tilde{C}(t,t')$ usually correspond to  eigenvectors which alternate in
sign (versus $t$) and so are not very  relevant to smooth fit functions.

As we have argued, with a small sample size $N$, the $D$ eigenvalues of 
the sample correlation matrix will be changed from their true values.
The largest relative effect will come when there are several true 
eigenvalues of similar  size - since those eigen directions mix fully. 
We find that the eigenvalue spectrum is densest at small eigenvalues.
This again leads to the conclusion that the smallest eigenvalues 
of the sample correlation matrix are the most strongly affected.

The SVD approach replaces the smallest (or zero) eigenvalues by  very
large values (since this corresponds to ignoring those terms in the
inverse).  We have explored this situation and found a less
discontinuous way to  modify the smallest eigenvalues $\lambda_i$ of
$\tilde{C}(t,t')$.  We keep the largest $E$ eigenvalues substantially
unchanged and then rearrange the remaining  smaller eigenvalues to avoid
extremely small ones.  Following from the motivation that like
eigenvalues mix most,  we propose to  redistribute the small eigenvalues
by replacing them  by their average to avoid very small values. 

After some experimentation, we propose the following explicit scheme to
replace the  $D$ eigenvalues $\lambda_i$ of the sample normalised
correlation  $\tilde{C}(t,t')$ (with convention $\lambda_i \ge
\lambda_{i+1}$) with  new eigenvalues $\lambda^{'}_i$:

 $$
  \lambda^{'}_{i} = K \ \hbox{Max} ( \lambda_i, \lambda_{min})
 $$
 $$
  \hbox{where} \ \ \lambda_{min}= {1 \over D-E} \sum_{i=E+1}^{D} \lambda_i
 \ \ 
  \hbox{and} \ \ K^{-1}= { 1 \over D} \sum_{i=1}^{D} 
\ \hbox{Max} ( \lambda_i, \lambda_{min})
 $$
The eigenvectors of $\tilde{C}(t,t')$ and thus of its inverse 
$\tilde{M}(t,t')$ are retained unchanged. Thus our procedure removes 
any very small eigenvalues of $\tilde{C}(t,t')$ and replaces them 
with the average of the $D-E$ smallest eigenvalues while retaining 
the property that the trace of $\tilde{C}(t,t')$ is $D$. The 
procedure also ensures a smooth eigenvalue distribution by 
allowing eigenvalues larger than this average to be retained.  
 
We tested this assignment with the same data as used above. Of course
for $E=D-1$, the method is equivalent to an exact inversion of the
sample correlation matrix.  For  $E > 2$, this method provides a stable 
model of  the  correlation matrix from the sample data. As $E$ is 
increased, the model reproduces  more exactly the sample 
correlation matrix - but at the expense of including unreasonable 
fluctuations if the sample size is too small. A compromise is to 
retain $E \approx \sqrt{N}$ exact eigenvalues when there are 
$N$ samples.

\section{Practical test of models}

Here we apply the various models described above to some typical  real
data. Since we need a large number of samples to give an  accurate data
set we chose some APE data~\cite{APE} on $\rho$  meson Green functions
$x(t)= \ < H(0) H(t) \! >$. Here  local-source and local-sink operators
$H$ are used for  creating and destroying a $\rho$ meson. The data set 
has $N=420$ samples of $x(t)$ with the clover fermionic action at
$K=.14190$ on  a $18^3\times 64$ lattice at $\beta=6.2$.

The data sample is large enough to allow a full correlated fit to  the
observed $x(t)$. An acceptable fit ($\chi^2/{\rm d.o.f.} 
=11.11/9$) to $x(t)$ with one exponential (actually a cosh is used) is
found for the $t$-range 14 to 24. This 2-parameter fit corresponds to 
requiring a plateau in the effective mass. The eigenvalues of the
normalised correlation  matrix for this data set are shown in fig~2 by
the continuous line.

Here our intention  is not to obtain the most precise values of
$m_{\rho}$ but to  illustrate the stability of various fitting
prescriptions. Thus we take this full data sample as the  true result
and explore fits using smaller subsets of the 420 data. For example we
take 14 blocks of 30 data. For each such set  of 30, we perform a fit to
the same $t$-range.  The intention is to  check whether the $\chi^2/{\rm
d.o.f.}$ is similar to the true value  from the full data set.  As well
as the average value of  $\chi^2/{\rm d.o.f.}$, one may study its
distribution so that confidence levels can  be extracted. We do not
pursue this here. 

A valid criterion for goodness of fit is of importance. The usual 
method is to use   $\chi^2$  to decide if a fit  is acceptable over a
given $t$-range. Thus an erroneous $\chi^2$ value  will lead to an
increase or decrease in the $t$-range chosen as acceptable. This in turn
will bias the fitted parameters such as $m_{\rho}$. For example, an 
uncorrelated fit will yield much lower $\chi^2$ which will then suggest
lower  $t$ values being included. This will tend to increase $m_{\rho}$ 
since the effective mass is a decreasing function of $t$ in this case.

Returning now to the comparison: fig~3 shows the results of the average 
of 14 fits to different samples of size $N=30$. The full  correlated fit
($n=11$) has a higher $\chi^2$ on average by 40\% than  the `true'
result. This  is entirely expected from the factor $1+(D+1)/N=1.39$
increase in  $\chi^2$ predicted~\cite{cmfit}. The eigenvalues from one
such sample fit to  $N=30$ are shown in fig~2. Here the phenomenon of
very small  eigenvalues appearing for small sample size is clear. Thus a
direct  use of a correlated fit to $N=30$ samples is likely to be
biassed  by those spurious small eigenvalues. We consider now some of 
the models introduced above to counter this while retaining a 
reasonable estimate of the goodness of fit.

As shown in fig~3, the  uncorrelated fit  has a very much reduced
$\chi^2$ as expected.  The  1- and 2-exponential models of section 4 do
much better in estimating  $\chi^2$. The eigenvalue  smoothing model of
section 5 also does well when 4 to 8 eigenvalues  are retained exactly
rather than all eigenvalues ($n=11$). Note that this is indeed 
consistent with our previous estimate that approximately $\sqrt{N}$ 
exact eigenvalues can be relied on. The modifications to the eigenvalues
from the smoothing  model are shown in fig~2 for one sample and 6 exact
eigenvalues. Indeed the modification  does alter the eigenvalues from
the sample in the direction of the true values. 

Also shown in fig~3 are  the results using an SVD definition of the inverse
of $\tilde{C}(t,t')$  in which $n$ eigenvalues are retained exactly and
the remainder are  discarded. It is possible to estimate the  expected
value of $\chi^2$ in this approach~\cite{kilcup}, yielding $N (n-2) /
(N-n-2)$ which is quite close to the values in fig~3. Thus  a corrected
estimator of the goodnes of fit is essential in this approach. The
eigenvalue distribution is not smooth since the deletion of  eigenmodes 
is  equivalent  to replacing the $D-n$ smallest eigenvalues in fig~2 by
infinite values.

For the samples of $N=30$, the situation is that  the correlations 
appear stronger than the true distributions.   This can be understood
from the  earlier discussion of the eigenvalues of  $\tilde{C}(t,t')$.
The true correlation matrix $\tilde{C}$ has 11  eigenvalues  with values
in the range  0.00438 to 9.82 whereas from subsets of 30 samples the 
smallest eigenvalue fluctuates between 0.00052 to 0.002844.  These
reductions in the magnitude of the smallest eigenvalues correspond to
narrower probability distributions and hence stronger correlations. The
implications of this for the  goodness of fit have been discussed, but
we also need to check that the  fitting procedures do not upset the
fitted parameters themselves. For  our chosen $t$-range the correlated
fit to the full 420 samples  yields $m_{\rho}a=0.351(5)$. A range of
other fits to the  same full sample (ie uncorrelated, 5-diagonal,
smoothed, etc) give  essentially the same value of $m_{\rho}$. Thus it
is only the goodness of  fit that depends on the correlation model used.
For the fits to subsets of 30 samples, we find that  an uncorrelated fit
does give the same result when averaged over  the 14 independent blocks.
The various correlation models all give  a result somewhat higher
($m_{\rho}a \approx 0.358$). Thus  we have some evidence that the 
uncorrelated fit is the most stable for determining the fit 
parameters when the sample size is reduced.

\section{Conclusion}

Data appropriate to hadron propagation in lattice gauge theory 
calculations are very strongly correlated. A full correlated fit 
to such data can be biassed unless the sample size ($N$) is 
sufficiently large compared with the number of data points ($D$). 
The main effect is the appearance of  spurious small eigenvalues 
of the correlation matrix. These increase the correlation in 
the sample. This increases $\chi^2$ by a factor of $(N-1)/(N-D-2)$.
It can also bias the fit parameters.

Here we propose models which can ameliorate this. Such models  depend,
to some extent, on an understanding of the expected  form of the
correlation matrix.  Thus such methods are not  completely general. For
the applications considered here, we  find two promising avenues. One is
to require that the  normalised correlation matrix has a tri-diagonal 
or 5-diagonal inverse. The other is to average the smallest  eigenvalues
so that spurious small values are removed. Both  of these models give
reasonable estimates of the goodness  of fit even with quite small
sample size. The goodness of  fit is important because it determines the
range of data  (eg. the $t$-range) to be fitted. This then will
influence the  fitted parameters in turn.

For the actual best determination of the fitted parameters for 
a small sample, we find that an uncorrelated fit is stable and thus 
an attractive proposition.

In fitting to hadron Green functions, it is  preferable  to
make a simultaneous fit to several operators.  For example 
smeared or fuzzed operators can be used at either sink or  source as
well as local ones. The observables for different  operators will be
correlated in general. We have discussed the  correlation in $t$, but
this correlation among different operators  will need  somewhat
different models. The most attractive route is  to use the method of
smoothing the eigenvalues of the correlation  matrix which is now
treated as a $PT \times PT$ matrix if there  are $T$ $t$-values and $P$
operators.

\section{Acknowledgements}

We thank our colleagues in the UKQCD Collaboration for 
helpful discussions and for allowing us access to their full data 
set. We also thank the APE Collaboration for making full data 
available to us. 

\appendix

\section{Appendix}

Consider the matrix given by
 $$
 \tilde{C}(t,t')={  (1-{\alpha_2}^2) {\alpha_1}^{|t-t'|+1}
                  - (1-{\alpha_1}^2) {\alpha_2}^{|t-t'|+1} 
  \over (\alpha_1-\alpha_2) (1+{\alpha_1} {\alpha_2})   }
 $$  
Its inverse $\tilde{M}(t,t')$ is 5-diagonal and 
is given by the following expressions 
in terms of
$p_1=\tilde{C}(t,t\pm 1)$ and $p_2=\tilde{C}(t, t \pm 2)$ with
$    d=1-2 {p_1}^2 + 2 {p_1}^2 p_2 - {p_2}^2$ 

 $$
 \tilde{M}(t,t \pm 1)=-{p_1(1-p_2)^2 \over d (1-p_1^2) } \ , \
 \tilde{M}(t,t \pm 2)={p_1^2-p_2  \over d}\ , \
 \tilde{M}(t,t)=1-2 p_1 \tilde{M}(t,t \pm 1) -2 p_2 \tilde{M}(t,t \pm 2)  \ , \ 
 $$
 $$
 \tilde{M}(t_m,t_m)={1-p_1^2 \over d} \ , \ 
 \tilde{M}(t_n,t_n)={1-p_1^2 \over d}- p_1 \tilde{M}(t,t \pm 1)	\ , \  
  \tilde{M}(t_m,t_m \pm 1 )=-{p_1 (1-p_2) \over d }.
 $$
where $t_m$ is the maximum or minimum value of $t$ and 
$t_n$ is the next to maximum or minimum.

\newpage

\begin{figure}[h]
\centerline{\psfig{figure=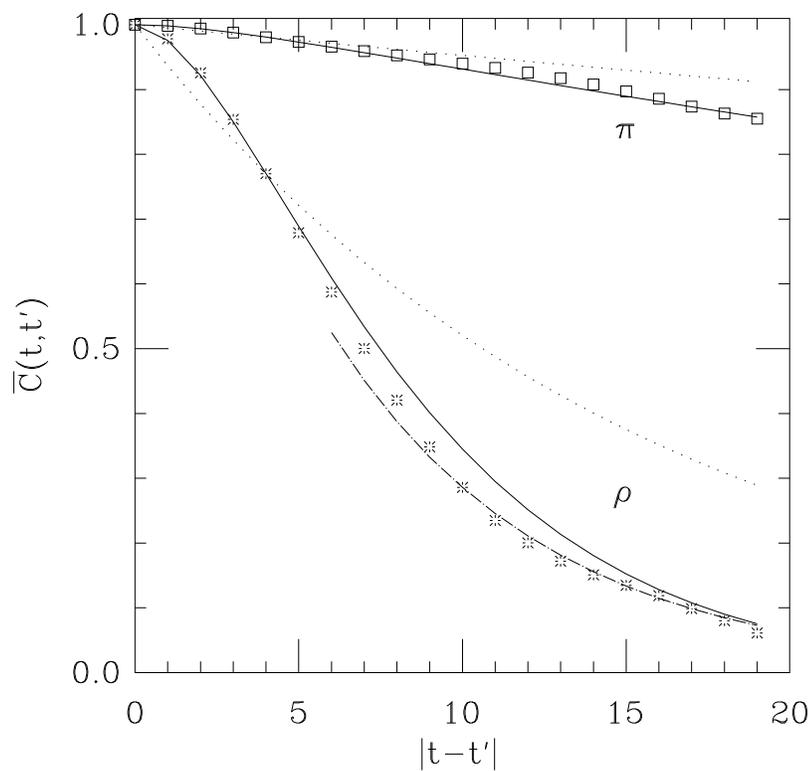,height=6in}}
\caption{ The normalised correlation $\tilde{C}(t,t')$ versus 
the relative $t$ difference for pion and rho Green functions from 
the data of section 3.
The continuous line is the 2-exponential model, the dotted line 
is one exponential and the dot-dashed curve is the expected 
behaviour from theoretical analysis of the 4 point function.
}
\end{figure}

\begin{figure}[h]
\centerline{\psfig{figure=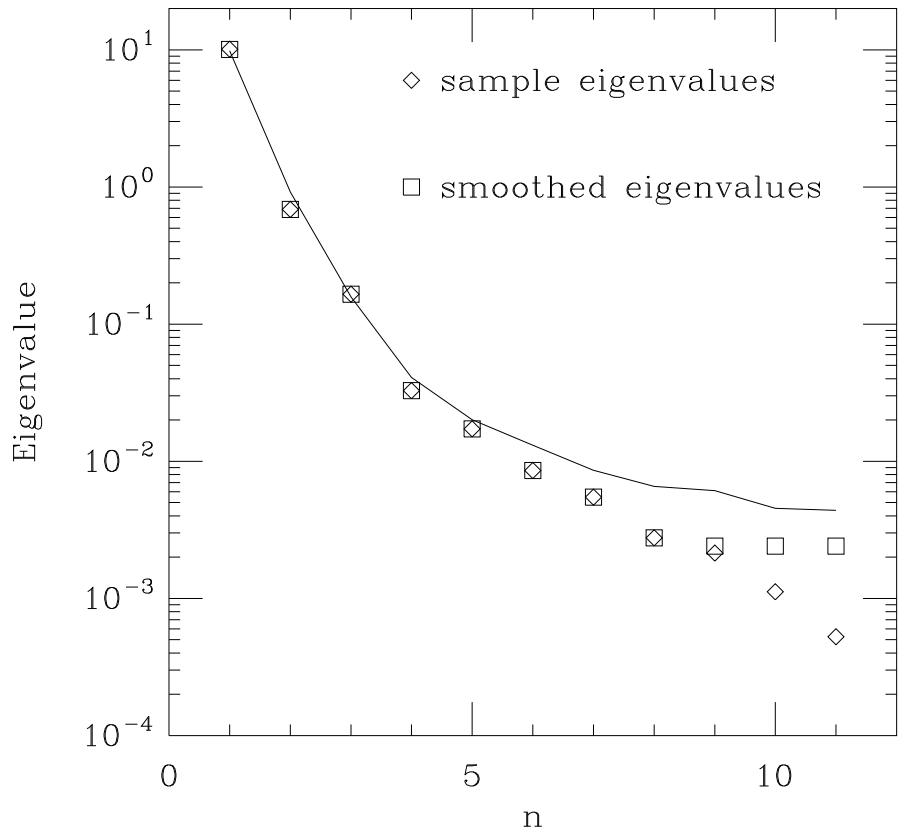,height=6in}}
\caption{ The eigenvalues of the $11 \times 11$ 
normalised correlation $\tilde{C}(t,t')$ for the data described  
in section 6. The full line is the 
result from $N=420$ configurations. The sample values (diamonds) are
from a subset of $N=30$. The squares represent the smoothed 
prescription of section 5 with 6 exact eigenvalues.
}
\end{figure}

\begin{figure}[h]
\centerline{\psfig{figure=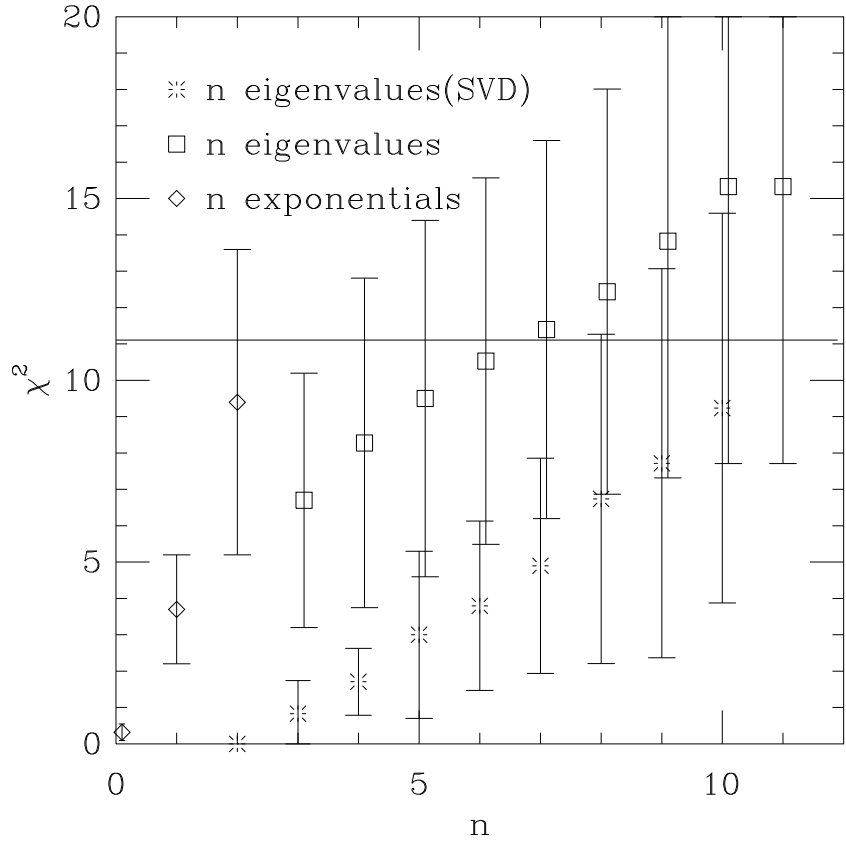,height=6in}}
\caption{ The $\chi^2$ for the $P=2$ parameter correlated fit to 
$D=11$ data. The full line is the result for the correlated fit to 
 the full data set of $N=420$. The expected $\chi^2$ is given by the
number of degrees  of freedom $D-P=9$. The  same fit was made to samples
of $N=30$ data. The average and standard deviation of 14 such
independent fits are  shown.  The uncorrelated, 1-exponential, and
2-exponential models  of section 4 are shown by diamonds. The eigenvalue
smoothed models  of section 5 are shown by squares. For comparison an
SVD-truncated model  is also shown (bursts). 
 }
\end{figure}

\end{document}